# CLASSICAL ANALOG OF SPIN

# IN 7 AND IN 10 DIMENSIONS

**Evangelos Chaliasos**

365 Thebes Street

GR-12241 Aegaleo

Athens, Greece

*Abstract*

Extra dimensions are introduced: 3 in Classical Mechanics and 6 in Relativistic Mechanics, which represent *orientations,* resulting from *rotations,* of a particle, described by *quaternions,* and leading to a **7**-dimensional, respectively **10**-dimensional, world. The concept of *spin* is thus defined in both cases, which results to be *different* from angular momentum. It leads to the known concept of Quantum Mechanical spin in both cases. In particular, application of these new ideas to (Relativistic) Quantum Mechanics, leads to quantization of energy levels in *free* electrons, which could be in principle detectable via their spectrum, confirming the theory.

# 1. Introduction

It is very well known the old dispute in Philosophy between *idealism* and *matterialism*. Namely, on the one hand idealism claims that the only true reality is the world of ideas, studied by Mathematics. The physical world then, studied by physics, is merely a kind of reflection of the world of ideas. On the other hand, matterialism claims that the only true reality is the physical world. The world of ideas is then obtained by merely a deduction from the physical world. In any case both agree in stating that there is a correspondence between these two worlds and this is which we are interested in concerning the pressent work.

In view of the above complete equivalence between these two worlds, it does not matter from which one will start in order to arrive at the other. In the present context, I think it is more natural and convenient to start from the physical world of Physics in order to go to the world of ideas of Mathematics. Namely we will consider the notion of a *point* in Physics to construct the concept of *point* in Mathematics.

As we all know our world or our (physical) *space* is, setting Relativity aside as a first step, *three-dimensional* (setting time apart at present). This is the same as stating that the "position" of a point is determined by *three* numbers, the three (spatial) coordinates. The point is thus considered to have no "magnitude" in itself, visualized as being the limiting object, taken from a real object, having a finite volume, if it is shrunk to zero (volume).

But we also know that a finite object possesses *orientation,* with respect to the coordinate axes, in addition to *position.* If we srink the object, keeping it "similar" (*homeothetous)* to itself at the same time, it will obviously retain its *orientation.* It seems to me as *paradox* the fact that when we arrive at the *limit* of a *point,* the orientation *disappears*. It is evidently more reasonable to consider that the orientation "survives" in the limit. And thus to adopt the concept of an *oriented point,* and the associated *orientation space* at a given *position.*

But the set of all orientations at a given position obviously results from an arbitrary orientation by performing all possible *rotations* of it. On the other hand each possible rotation can be analyzed to *three* rotations, each of them taking place on each of the *three* (3) coordinate planes (xy, yz, zx) Thus, the *orientation space* has to be *three*-dimensional. This conclusion can also result from the fact that the rotation of an extended body is specified by its *three* (3) Eulerian angles. Now mathematically the concept of such a rotation is described by a *quaternion*. Thus we are led to describe the *orientations* of a point by *quaternions*.

In this way we conclude that we have to add *three* numbers (related with orientation) to the usual *three* coordinates (related to position) in order to describe a point. Thus, if we add the time, we are led to the conclusion that our world is *seven-dimensional!*

In Relativity, as we all know, the *position* of a point in Minkowski space is determined by *four* (4) coordinates. This is the case because *space* and *time* are now unified. In the same manner as before we have to associate now to the *four*-dimensional space-time and to each *position* the set of all *orientations* of a *four*-dimensional coordinate system. This is evidently the same with the set of all *rotations* of this four-dimensional coordinate system. These are *six*. Three of them express rotations of the space alone as before. The other three, expressing pure *boosts* (Lorentz transformations), are specified by "rotations" on each of the planes xt, yt, zt. Thus we arrive at a new conception, the one of a (4+3+3=10) *ten-dimensional world!* The present paper is but a first attempt to grasp this world.

Note that extra dimensions are already in use in modern quantum field theory. They are (ten in number, too) involved mainly in the theory of superstrings, through which they were introduced. But the nature of these extra dimensions is not specified, being something complettely irrelevant to the ideas exposed above.

## 2. Classical Mechanics in 7 dimensions

*i) Introduction of quaternions*

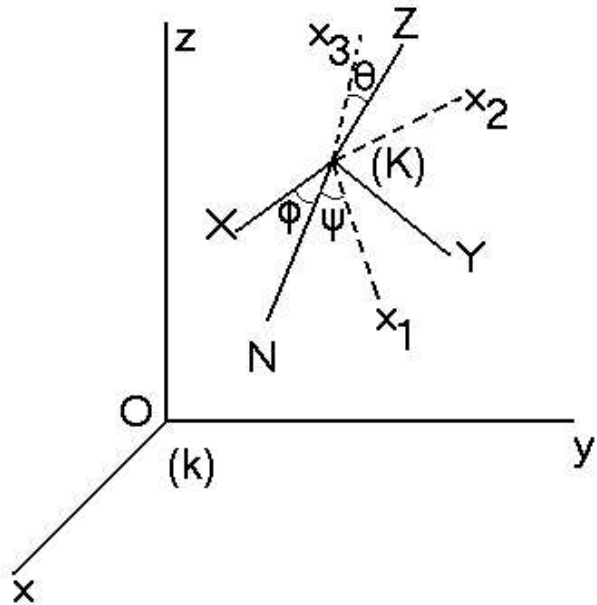

*Figure 1 Eulerian angles (θ, φ, ψ) of "inner" orientation (x1, x2, x3) of point A relative to system K (X, Y, Z). AN is the section of plane (X, Y) with plane (x1, x2). The system k (x, y, z) is the usual position frame.*

Consider a cartesian and rectangular coordinate system (k) in space with axes (x, y, z). to an arbitrary point A its three coordinates x, y, z are attributed. According to the established ideas these three coordinates are enough in order to completely specify the point. We accept that these three coordinatess are not enough, and that each point A possesses an "orientation" described by the three Eulerian angles, for example, (θ, φ, ψ) relative to a cartesian and rectangular coordinate system (K) with origin at A and axes (X, Y, Z), as in fig. 1 (see also Landau & Lifshitz, 1976, for the analog with respect to a rigid body). Then the (generalized) coordinates qi of A are *six,* namely (x, y, z; θ, φ, ψ) and the physical space is six-dimensional, or the world is *seven*-dimensional, if we add the time t. In order for the orientations (θ, φ, ψ) of points at different positions (x, y, z) to be comparable, we take the K-systems at different positions to have their axes (X, Y, Z) parallel to one another, but not necessarilly parallel to (x, y, z) (exept as a special case).

The displacement of a position A relative to O is **OA**. If the displacement of a position B relative to A is **AB**, then the displacement of B relative to O is given by **OA+AB.** In other words the "composition" of positions is given by the usual vector addition. But concerning the "composition" of "orientations" at A, these orientations must be viewed as "rotations" of A from (X, Y, Z) to (x1, x2, x3). Then, in order to have the rule for the "composition" of these "rotations", the latter must be described in terms of *quaternions*. Then the "composition" of "orientations" is given by the usual composition of quaternions. Supposse that the orientation (θ, φ, ψ) of A is thus given by the quaternion

$$R = \cos(\theta/2) - i\sin(\theta/2)\left(\sigma_x \cos\alpha + \sigma_y \cos\beta + \sigma_z \cos\gamma\right) \quad (1)$$

or

$$R_{\vec{n}}(\theta) = e^{-i(\theta/2)\vec{\sigma}\cdot\vec{n}} \quad (2)$$

(see Misner, Thorne & Wheeler, 1973). The quaternion (1) is described by the four parameters θ and (cosα, cosβ, cosγ). But in the four-dimensional space of quaternions we are restricted on a three-dimensional hypersurface only, given that

$$\cos^2\alpha + \cos^2\beta + \cos^2\gamma = 1. \quad (3)$$

This hyper-surface can be described by the three components of the vector **θ,** whose magnitude is θ and direction is **n**. Thus each quaternion R can be described by a vector **θ**R in (K), with the only difference that with regard to composition we have to adopt the composition of quaternions rather than the composition of vectors.

*ii) The Lagrangian of a particle and its moment of inertia*

Suppose now that we have a free *material* point A fixed in position space. Then its Lagrangian will not obviously depend on its position coordinates and on their corresponding velocities. The Lagrangian will not also depend on time, because of the homogeneity of time. Futhermore, we assume that the orientation space is homogeneous, reflecting the isotropy of the position space. Then the Lagrangian will not depend on the orientation (x1, x2, x3) relative to (X, Y, Z), depending only on the rate of change of orientation, that is on the angular velocity **ω** of (x1, x2, x3) relative to (X, Y, Z). Because also all orientations of (x1, x2, x3) are equivalent, the Lagrangian will not depend on the direction of **ω** relative to (X, Y, Z), but only on its magnitude $\omega^2$ and, perhaps, on its direction relative to (x1, x2, x3), expressed by its three direction cosines α, β, γ, being the components of a unit vector **n.** Thus:

$$L = L(\vec{n}, \omega^2). \quad (4)$$

Suppose now that, leaving the direction **n** unaffected, we increase ω by dω to ω´ = ω + dω. Since the equations of motion retain the same form regardless of the substitution ω → ω´, it is evident that the "new" Lagrangian will be

$$L' = L(\vec{n}, \omega'^2).$$

But because exactly both L and L´ must lead to equations of motion of the same form, they must differ only by a total time derivative of a function of orientation coordinates (see Landau & Lifshitz, 1976). Expanding L´ to first order in dω, we have

$$L' = L(\vec{n}, \omega^2 + 2\omega d\omega) = L + \frac{\partial L}{\partial \omega^2} \cdot 2\omega d\omega. \quad (5)$$

But in order for the last term in eqn. (5) to be a total time derivative of a function of orientation coordinates, it must be linear in the rate of change of orientation. This can happen only if $\partial L/\partial \omega^2$ is independent of $\omega^2$, L thus being "proportional" to $\omega^2$. Therefore, we can write

$$L = (1/2)I(\vec{n})\omega^2, \quad (6)$$

Where the "constant" of proportionality I may depend in general on **n**, and it is called the "moment of inertia" about the **n** axis. As in the case of the mass of the particle, it can be shown that I is always positive (see Landau & Lifshitz, 1976).

Now, suppose that we fix $\omega^2$ in L and we change α, β, γ - which we call αi (i = 1, 2, 3) – from αi to α´i = αi + dαi. Then L will be changed from L = L(αi, $\omega^2$) to L´ = L(α´i, $\omega^2$). Expanding it to first order we find

$$L' = L + \sum_i \frac{\partial L}{\partial \alpha_i} d\alpha_i. \quad (7)$$

Because with this change L must retain the same form, it must transform in the same manner with αi. But, because it is about a rotation, dαi are expressed as linear combinations of αi with infinitesimal coefficients. Because then $\Sigma(\partial L/\partial \alpha i)d\alpha i$ has to be expressed in the same way, it follows that $\partial L/\partial \alpha i$ must be linear in αj´s and, therefore, that L is a quadratic form with respect to the direction cosines. Thus, extracting (½)$\omega^2$ as a common factor, we can write

$$I(\alpha, \beta, \gamma) = I_{xx}\alpha^2 + I_{yy}\beta^2 + I_{zz}\gamma^2 + 2I_{xy}\alpha\beta + 2I_{yz}\beta\gamma + 2I_{zx}\gamma\alpha. \quad (8)$$

We call the elements of the symmetric matrix

$$II = \begin{pmatrix} I_{xx}, I_{xy}, I_{xz} \\ I_{yx}, I_{yy}, I_{yz} \\ I_{zx}, I_{zy}, I_{zz} \end{pmatrix} \quad (9)$$

the components of the "moment of inertia tensor" Iij relative to (x1, x2, x3). It remains of course to be shown that they are in fact the components of a tensor. For this see Goldstein, 1968, p. 147.

To find the critical points of the quadratic form (8), we have to diagonalize the matrix (9). We then have

$$II = \begin{pmatrix} I_1, 0, 0 \\ 0, I_2, 0 \\ 0, 0, I_3 \end{pmatrix} \quad (10)$$

with the eigenvalues positive. We assume that (in general) I1 < I2 < I3. From the three "principal moments of inertia" I1, I2, I3 we have that I1 is the (absolute) minimum, I3 is the (absolute) maximum, and I2 is a saddle point of the quadratic form I($\alpha$, $\beta$, $\gamma$). The three eigenvectors are normal to one another and define the three principal axes of inertia. We will always, from now on, take as orientation (x1, x2, x3) of the particle the one defined by the three principal axes of inertia. Then the quadratic form (8) takes the form

$$I(\alpha, \beta, \gamma) = I_1 \alpha^2 + I_2 \beta^2 + I_3 \gamma^2, \quad (11)$$

so that (6) becomes

$$L = (1/2) I_1 \omega_1^2 + (1/2) I_2 \omega_2^2 + (1/2) I_3 \omega_3^2, \quad (12)$$

where $\omega_1$, $\omega_2$, $\omega_3$ are the three components of $\boldsymbol{\omega}$ along the axes x1, x2, x3 respectively.

*iii) Classical spin and its conservation*

We define the *spin* **s** of the particle as

$$\vec{s} := II \vec{\omega}, \quad (13)$$

so that, in the "body" system,

$$s_1 = I_1\omega_1, \quad s_2 = I_2\omega_2, \quad s_3 = I_3\omega_3. \qquad (14)$$

We will prove that **s** is conserved in the "space" system. From the Euler equations (resulting from Lagrange's equations) we have in the body system

$$I_1\dot{\omega}_1 + \omega_2\omega_3(I_3 - I_2) = 0, \qquad (15)$$

or

$$I_1\dot{\omega}_1 + (\omega_2 I_3\omega_3 - \omega_3 I_2\omega_2) = 0, \qquad (16)$$

or

$$I_1\dot{\omega}_1 + (\vec{\omega} \times \vec{s})_1 = 0. \qquad (17)$$

But

$$\left(\frac{d}{dt}\right)_{body} + \vec{\omega} \times \quad = \left(\frac{d}{dt}\right)_{space} \qquad (18)$$

(see Goldstein, 1968). Thus

$$ds_1/dt = 0, \qquad (19)$$

where d/dt means space differentiation, and similarly for the other components. Therefore

$$\frac{d\vec{s}}{dt} = 0, \qquad (20)$$

QED.

*iv) Interaction of position with orientation*

If the material point is fixed in orientation space, then

$$L = (1/2)mv^2 \qquad (21)$$

(see Landau & Lifshitz, 1976). In the general case thus

$$L = L_1 + L_2 + L_{12}, \qquad (22)$$

where L1 is due to position, that is

$$L_1 = (1/2)mv^2, \qquad (23)$$

L2 is due to orientation, that is

$$L_2 = (1/2)I\omega^2, \qquad (24)$$

and L12 is due to the interaction, in general, between position and orientation. In order for L to give L1 and L2 in the special cases that the particle is fixed in orientation and in position space respectively, **v** and **ω** must be common factors in L12, so that

$$L_{12} = A\vec{v} \cdot \vec{\omega}, \qquad (25)$$

with A a constant (independent of position and orientation). Then the Lagrange equations give

$$\partial L/\partial \vec{v} = const., \qquad (26)$$

so that

$$m\vec{v} + A\vec{\omega} = const.. \qquad (27)$$

For **ω** = 0, we have the law of inertia. In general, if **ω** precesses about the spin axis, then **v** changes, giving rise to motion on a helix. Because this helix must be very close to a straight line, A must be

very small. Also, the Euler equations now become

$$I_1\dot{\omega}_1 - \omega_2\omega_3(I_2 - I_3) = -A\dot{v}_1 + A(\vec{v}\times\vec{\omega})_1, etc., \quad (28)$$

so that

$$d\vec{s}/dt = -A\dot{\vec{v}} + A\vec{v}\times\vec{\omega}. \quad (29)$$

But

$$\dot{\vec{v}} = d\vec{v}/dt - \vec{\omega}\times\vec{v}, \quad (30)$$

so that finally

$$\frac{d\vec{s}}{dt} = -A\frac{d\vec{v}}{dt}, \quad (31)$$

or, from (27),

$$\frac{d\vec{s}}{dt} = \frac{A^2}{m}\frac{d\vec{\omega}}{dt}. \quad (32)$$

And, because of (13), we have finally the matrix differential equations

$$\frac{d\vec{\omega}}{dt} = \frac{m}{A^2}\frac{d}{dt}\{II(t)\vec{\omega}\}. \quad (33)$$

This is an equation for the "space" components. Now express ω as a function of the Euler angles (see Goldstein, 1968). Also, if the matrix B transforms the coordinates from the "space" to the "body" set of axes, then II will be given by the following similarity transformation

$$II = B^{-1}\begin{pmatrix} I_1, 0, 0 \\ 0, I_2, 0 \\ 0, 0, I_3 \end{pmatrix} B. \quad (34)$$

Expressing B and B^(-1) as functions of the Euler angles (see Goldstein, 1968), we can have from (33) three differential equations for the three Euler angles. After finding them, we will be able to determine **ω** as a function of t, and then, from (27), also **v** as a function of t.

As a first approximation, we can take A = 0 in (25). Then the motion of the particle in position space will be described as usually, taking the known law of inertia. The motion in orientation space will be given in the same way as the usual motion of a rigid body with one point fixed.

3. **Quantum Mechanics in 7 dimensions**

*i) Generalized momenta & the Hamiltonian in terms of the Eulerian angles*

Let us see what happens with respect to Quantum Mechanics. We will consider motion of a particle in orientation space, setting L = L2. Considering also a "spherical" particle, we will have

$$L = (1/2)I(\omega_1^2 + \omega_2^2 + \omega_3^2). \quad (35)$$

Expressing the components of **ω** in terms of the Euler angles (see Goldstein, 1968), we find

$$L = (1/2)I(\dot{\phi}^2 + \dot{\theta}^2 + \dot{\psi}^2 + 2\dot{\phi}\dot{\psi}\cos\theta). \quad (36)$$

We have thus

$$p_\phi \equiv \frac{\partial L}{\partial \dot{\phi}} = I(\dot{\phi} + \dot{\psi}\cos\theta), \quad (38)$$

$$p_\theta \equiv \frac{\partial L}{\partial \dot{\theta}} = I\dot{\theta}, \quad (38)$$

$$p_\psi \equiv \frac{\partial L}{\partial \dot{\psi}} = I(\dot{\psi} + \dot{\phi}\cos\theta). \quad (39)$$

Considering motion only through the angle φ (setting θ´ = ψ´ = 0), so that pφ = Iφ´, and doing the transition to Quantum Mechanics via the substitution

$$p_\phi \rightarrow \hat{p}_\phi = -i\hbar \frac{\partial}{\partial \phi},$$

we have for pφ the eigenvalue equation

$$\hat{p}_\phi /\psi> = p_\phi /\psi>.$$

The solution of this equation for /ψ> gives

$$/\psi> = e^{\frac{i}{\hbar} p_\phi \phi}.$$

Changing the argument by nπ, for a change in φ by 2π, gives pφ = (n/2)ℏ and thus changes the sign of /ψ> for n odd, while leaves /ψ> unaltered for n even. And since, as we will see in the sequel, the square of the spin has the eigenvalue

$$\lambda = \tfrac{1}{2}(\tfrac{1}{2}+1)\hbar^2,$$

we conclude that its projection pφ on the z-axis cannot be greater than ℏ√[½(½+1)], and thus

$$-\hbar\sqrt{3}/2 \leq p_\phi \leq \hbar\sqrt{3}/2.$$

And, since

$$p_\phi = (n/2)\hbar,$$

we find finally that

$$p_\phi = \pm\tfrac{1}{2}\hbar,$$

which means that the direction of spin is quantized. The quantization of direction gives

$$\cos\theta = \frac{\pm 1/2}{\sqrt{3}/2} = \pm\frac{1}{\sqrt{3}} \neq \pm 1,$$

or

$$1-\alpha^2 \equiv \sin^2\theta = 1-\frac{1/4}{3/4} = \frac{2}{3} \neq 0.$$

Thus, we conclude that the spin precesses around the z-axis.

From (37) and (39) we find

$$\dot\phi = \frac{p_\phi}{I\sin^2\theta} - \frac{p_\psi \cos\theta}{I\sin^2\theta} \qquad (40)$$

and

$$\dot\psi = \frac{p_\psi}{I\sin^2\theta} - \frac{p_\phi \cos\theta}{I\sin^2\theta}. \qquad (41)$$

We have thus the generalized velocities in terms of the generalized momenta, so that we find for the Hamiltonian finally

$$H \equiv \sum_i p_i q_i - L = \frac{p_\theta^2}{2I} + \frac{p_\phi^2 + p_\psi^2 - 2p_\phi p_\psi \cos\theta}{2I\sin^2\theta}. \qquad (42)$$

For constant θ, setting cosθ ≡ α, we have pθ = 0 and the Hamiltonian simplifies to

$$H = \frac{1}{2I(1-\alpha^2)}\{p_\phi^2 + p_\psi^2 - 2p_\phi p_\psi \cos\theta\}. \qquad (43)$$

*ii) Quantum Mechanics using the Eulerian angles*

Doing the transition to Quantum Mechanics by means of the formulae

$$\phi \to -i\hbar \frac{\partial}{\partial \phi} \quad \& \quad \psi \to -i\hbar \frac{\partial}{\partial \psi}, \quad (44)$$

the Schrodinger equation

$$\hat{H}|\psi> = E|\psi> \quad (45)$$

takes the form

$$-\hbar^2 \left( \frac{\partial^2}{\partial \phi^2} + \frac{\partial^2}{\partial \psi^2} - 2\alpha \frac{\partial^2}{\partial \phi \partial \psi} \right) |\psi> = 2I(1-\alpha^2)E|\psi>. \quad (46)$$

Even simpler: set $\psi$ = constant. Then (46) gives

$$\psi''(x) + \frac{2I}{\hbar^2}(1-\alpha^2)E\psi(x) = 0, \quad (47)$$

where we have renamed the Euler angle φ to x. If we set

$$\frac{2I}{\hbar^2}(1-\alpha^2)E \equiv k^2, \quad (48)$$

then the characteristic equation for (47) is

$$\sigma^2 + k^2 = 0. \quad (49)$$

(see Kappos, 1965). We have from (49) σ = ± ik, so that

$$\psi(x) = C_1 e^{ikx} + C_2 e^{-ikx}, \quad (50)$$

or

$$\psi(x) = A\cos kx + B\sin kx. \quad (51)$$

We thus obtain a harmonic solution. And because we demand that ψ(x + 2π) = ± ψ(x), we must have again k = n, or k = n/2, so that ψ(x + 2π) = - ψ(x) for n odd, while ψ(x + 2π) = ψ(x) for n even.

Now, from eqn. (48), we get

$$\frac{2I}{\hbar^2}\frac{2}{3}E = \frac{n^2}{4},$$

so that the energy E associated with the z-axis is given by

$$E = \frac{3n^2}{16I}\hbar^2.$$

To determine n, we compare this result with the one taken classically if we substitute the (quantized) pφ (= ± ½ ℏ, see above) in (the classical) eqn. (43) (with pψ = 0), which gives

$$E = \frac{3}{16I}\hbar^2.$$

We thus find that n = ± 1 (or k = ± ½). We see that the energy is the same in either case of the z-spin eigenvalue ± ½ ℏ, as expected.

**4. Spinors in orientation space**

*i) Introduction of spinors*

We have the relation (13)

$$\vec{s} = II\vec{\omega}. \quad (54)$$

We convert it to a spinor relation. We have on the one hand

$$S^{A\dot{U}} = s^{\mu}\sigma_{\mu}{}^{A\dot{U}} = \begin{pmatrix} s_3 & s_1 - is_2 \\ s_1 + is_2 & -s_3 \end{pmatrix} \quad (55)$$

On the other hand

$$\Omega^{A\dot{U}} = \omega^{\mu}\sigma_{\mu}{}^{A\dot{U}} \quad (56)$$

and

$$I^{A\dot{U}}{}_{B\dot{V}} = \sigma_{\mu}{}^{A\dot{U}}\sigma^{\nu}{}_{B\dot{V}}\Pi^{\mu}{}_{\nu}, \quad (57)$$

so that

$$I^{A\dot{U}}{}_{B\dot{V}}\Omega^{B\dot{V}} = \sigma_{\mu}{}^{A\dot{U}}\sigma^{\nu}{}_{B\dot{V}}\sigma_{\rho}{}^{B\dot{V}}\Pi^{\mu}{}_{\nu}\omega^{\rho} =$$
$$= -2\sigma_{\mu}{}^{A\dot{U}}\delta^{\nu}{}_{\rho}\Pi^{\mu}{}_{\nu}\omega^{\rho} = -2\sigma_{\mu}{}^{A\dot{U}}I^{\mu}{}_{\nu}\omega^{\nu}. \quad (58)$$

Thus, we find finally

$$I^{A\dot{U}}{}_{B\dot{V}}\Omega^{B\dot{V}} = -2\begin{pmatrix} I_3\omega_3 & I_1\omega_1 - iI_2\omega_2 \\ I_1\omega_1 + iI_2\omega_2 & -I_3\omega_3 \end{pmatrix} \quad (59)$$

so that

$$S^{A\dot{U}} = -\tfrac{1}{2}I^{A\dot{U}}{}_{B\dot{V}}\Omega^{B\dot{V}}. \quad (60)$$

*ii) Determination of the angular velocity*

We have to express ω1, ω2, ω3 in terms of the quaternion parameters θ; α, β, γ and their derivatives.

If the initial orientation is given by the quaternion

$$R_1 = \cos(\theta/2) + i\sin(\theta/2)\vec{\sigma}\cdot\vec{n}, \quad (61)$$

where **σ** the vector with components the Pauli matrices and **n** = (α, β, γ), and we perform the elementary rotation given by the quaternion

$$R(\vec{\omega}dt) = 1 + (i/2)\vec{\sigma}\cdot\vec{\omega}dt \qquad (62)$$

(see Misnere, Thorn & Wheeler, 1973), then the final orientation will be given by the quaternion

$$R_2 = R(\vec{\omega}dt) \circ R_1. \qquad (63)$$

We find to the first order

$$R_2 = \cos(\theta/2) + i\sin(\theta/2)\vec{\sigma}\cdot\vec{n} + (i/2)\vec{\sigma}\cdot\vec{\omega}dt\cos(\theta/2) - $$
$$- (1/2)\sin(\theta/2)(\vec{\omega}\cdot\vec{n})dt - (i/2)\sin(\theta/2)(\vec{\omega}\times\vec{n})\cdot\vec{\sigma}dt. \qquad (64)$$

On the other hand the final orientation, given by the quaternion R2, will correspond to quaternion parameters $\theta' = \theta + \Delta\theta$ and $\mathbf{n}' = \mathbf{n} + \Delta\mathbf{n}$, so that expanding the formula

$$R_2 = \cos(\theta'/2) + i\sin(\theta'/2)\vec{\sigma}\cdot\vec{n}', \qquad (65)$$

giving R2, in Taylor series, we find to first order

$$R_2 = \cos(\theta/2) - \sin(\theta/2)(\Delta\theta/2) + i\{\sin(\theta/2) + \cos(\theta/2)(\Delta\theta/2)\}\vec{\sigma}\cdot(\vec{n} + \Delta\vec{n}). \qquad (66)$$

Equating real and imaginary parts between (66) and (64), we find the equations

$$(\vec{\omega}\cdot\vec{n})dt = \Delta\theta \qquad (67)$$

and

$$½\cos(\theta/2)\vec{\sigma}\cdot\vec{\omega}dt - ½\sin(\theta/2)\vec{\sigma}\cdot(\vec{\omega}\times\vec{n})dt = $$
$$= ½\cos(\theta/2)\vec{\sigma}\cdot\vec{n}\Delta\theta + \sin(\theta/2)\vec{\sigma}\cdot\Delta\vec{n}. \qquad (68)$$

From (67) and (68) we find for the derivatives of the orientation (quaternion) parameters $\theta$ and $\mathbf{n}$

$$\dot{\theta} = \vec{\omega}\cdot\vec{n} \qquad (69)$$

and

$$\sin(\theta/2)\dot{\vec{n}} = \tfrac{1}{2}\cos(\theta/2)\vec{\omega} - \tfrac{1}{2}\sin(\theta/2)(\vec{\omega}\times\vec{n}) - \tfrac{1}{2}\cos(\theta/2)\vec{n}(\vec{\omega}\cdot\vec{n}). \quad (70)$$

We are interested in expressing **ω** in terms of the orientation parameters and their derivatives, in order to insert ω1, ω2, ω3 in (59). Thus, solving (69) for ω1 and inserting in (70) (one-component), we find (setting $c \equiv \cos(\theta/2)$ and $s \equiv \sin(\theta/2)$ for brevity)

$$c\left(\frac{\dot{\theta}}{\alpha} - \frac{\beta}{\alpha}\omega_2 - \frac{\gamma}{\alpha}\omega_3\right) - s(\omega_2\gamma - \omega_3\beta) =$$
$$= c\alpha\dot{\theta} + 2s\dot{\alpha} - c\dot{\theta}/\alpha, \quad (71)$$

from which we find ω2 as a function of ω3. Inseerting this into (69) we find

$$\omega_1 = \frac{\dot{\theta}}{\alpha} - \frac{\beta}{\alpha}\left(\frac{B}{A} + \frac{C}{A}\omega_3\right) - \frac{\gamma}{\alpha}\omega_3, \quad (72)$$

with A, B, C appropriate quantities, i.e. we find ω1 in terms of ω3. Solving also (69) for ω2 and inserting in (70) (two-component), we find

$$c\left(\frac{\dot{\theta}}{\beta} - \frac{\alpha}{\beta}\omega_1 - \frac{\gamma}{\beta}\omega_3\right) - s(\omega_3\alpha - \omega_1\gamma) = c\beta\dot{\theta} + s\dot{\beta}. \quad (73)$$

Solving (73) for ω1 we find it in terms of ω3. Equating this to ω1 from (72) we have an equation for ω3. After some manipulations we write it in the form

$$\{-\beta(c\gamma - s\beta\alpha)(-c\alpha + s\gamma\beta) - \gamma(-c\beta - s\gamma\alpha)(-c\alpha + s\gamma\beta) -$$
$$- \alpha(c\gamma + s\alpha\beta)(-c\beta - s\gamma\alpha)\}\omega_3 =$$
$$= -\dot{\theta}(-c\beta - s\gamma\alpha)(-c\alpha + s\gamma\beta) + \beta c\dot{\theta}(\alpha^2 - 1)(-c\alpha + s\gamma\beta) +$$
$$+ 2s\alpha\dot{\alpha}\beta(-c\alpha + s\gamma\beta) + c\dot{\theta}\alpha(\beta^2 - 1)(-c\beta - s\gamma\alpha) + 2s\beta\dot{\beta}\alpha(-c\beta - s\gamma\alpha). \quad (74)$$

We find after some manipulations

$$\{\quad\} = \alpha\beta\gamma. \quad (75)$$

For the right-hand side of (74) we find

$$\dot{\theta}\alpha\beta\gamma^2 + 2cs\alpha\beta\gamma\dot{\gamma} + 2s^2\alpha\beta\gamma(\dot{\alpha}\beta - \dot{\beta}\alpha), \quad (76)$$

where we used the identity

$$\alpha^2 + \beta^2 + \gamma^2 = 1, \quad (77)$$

resulting in

$$\alpha\dot{\alpha} + \beta\dot{\beta} + \gamma\dot{\gamma} = 0. \quad (78)$$

Thus, finally, we find from (74)

$$\omega_3 = \dot{\theta}\gamma + 2cs\dot{\gamma} + 2s^2(\dot{\alpha}\beta - \dot{\beta}\alpha) \quad (79)$$

or

$$\omega_3 = \dot{\theta}\gamma + \sin\theta\dot{\gamma} + (1-\cos\theta)\begin{vmatrix} \dot{\alpha} & \dot{\beta} \\ \alpha & \beta \end{vmatrix}. \quad (80)$$

In the same way, by cyclic permutation of the "indices", we find ω1 and ω2, with the final result

$$\vec{\omega} = \dot{\theta}\vec{n} + (\sin\theta)\dot{\vec{n}} + (1-\cos\theta)\dot{\vec{n}}\times\vec{n}. \quad (81)$$

Inserting (81) in (70), written in the form

$$c\vec{\omega} - s\vec{\omega}\times\vec{n} = c\dot{\theta}\vec{n} + 2s\dot{\vec{n}}, \quad (82)$$

or

$$(1+\cos\theta)\vec{\omega} - (\sin\theta)\vec{\omega}\times\vec{n} = (1+\cos\theta)\dot{\theta}\vec{n} + 2(\cos\theta)\dot{\vec{n}}, \quad (83)$$

and expanding the double cross-product appearing according to

$$(\dot{\vec{n}} \times \vec{n}) \times \vec{n} = -\dot{\vec{n}}, \quad (84)$$

since

$$\vec{n} \times \dot{\vec{n}} = 0, \quad (85)$$

we find indeed that (70) is satisfied.

*iii) The angular velocity continued*

If we use the variable $t \equiv \tan(\theta/2)$ instead of $\theta$, we have

$$\sin\theta = \frac{2t}{1+t^2}, \quad (86)$$

$$1 - \cos\theta = \frac{2t^2}{1+t^2}, \quad (87)$$

and, because then $\theta = 2\arctan t$,

$$\dot{\theta} = \frac{2\dot{t}}{1+t^2}. \quad (88)$$

Thus, (81) becomes

$$\vec{\omega} = \frac{2\dot{t}}{1+t^2}\vec{n} + \frac{2t}{1+t^2}\dot{\vec{n}} + \frac{2t^2}{1+t^2}\dot{\vec{n}} \times \vec{n}. \quad (89)$$

Now, instead of $\alpha, \beta, \gamma$ we use the variables (generalized coordinates)

$$x \equiv \alpha t, \quad y \equiv \beta t, \quad z \equiv \gamma t. \quad (90)$$

We have then

$$x^2 + y^2 + z^2 = t^2, \quad (91)$$

so that

$$t = \sqrt{x^2 + y^2 + z^2}, \quad (92)$$

and

$$\alpha = \frac{x}{\sqrt{x^2 + y^2 + z^2}}, \quad \beta = \frac{y}{\sqrt{x^2 + y^2 + z^2}}, \quad \gamma = \frac{z}{\sqrt{x^2 + y^2 + z^2}}. \quad (93)$$

We have thus

$$\vec{n} = \vec{t}/t, \quad (94)$$

and we find

$$\dot{\vec{n}} = \frac{\dot{\vec{t}}}{t} - \frac{\vec{t} \cdot \dot{\vec{t}}}{t^2} \frac{\vec{t}}{t}, \quad (95)$$

and

$$\dot{\vec{n}} \times \vec{n} = \frac{\dot{\vec{t}} \times \vec{t}}{t^2}. \quad (96)$$

Substituting then **n, n´ & n´× n** from (94), (95) & (96) respectively in (89) yields

$$\frac{1+t^2}{2} \vec{\omega} = \dot{\vec{t}} + \dot{\vec{t}} \times \vec{t}, \quad (97)$$

that is, for example,

Now, the Lagrangian for a free particle at rest in position space is given by (12)

$$\frac{1+x^2+y^2+z^2}{2}\omega_1 = \dot{x} + \begin{vmatrix} \dot{y} & \dot{z} \\ y & z \end{vmatrix}. \qquad (98)$$

$$L = \frac{1}{2}I_1\omega_1^2 + \frac{1}{2}I_2\omega_2^2 + \frac{1}{2}I_3\omega_3^2 \qquad (99)$$

in the "body" set of axes. Substituting in (99) ω1, ω2, ω3 from (97), we have

$$L = \frac{2I_1}{(1+t^2)^2}\left(\dot{x} + \begin{vmatrix} \dot{y} & \dot{z} \\ y & z \end{vmatrix}\right)^2 + \frac{2I_2}{(1+t^2)^2}\left(\dot{y} + \begin{vmatrix} \dot{z} & \dot{x} \\ z & x \end{vmatrix}\right)^2 + \frac{2I_3}{(1+t^2)^2}\left(\dot{z} + \begin{vmatrix} \dot{x} & \dot{y} \\ x & y \end{vmatrix}\right)^2. \qquad (100)$$

Thus we find for the generalized momenta px etc., for example,

$$p_x \equiv \frac{\partial L}{\partial \dot{x}} = \frac{4I_1}{(1+t^2)^2}\left(\dot{x} + \begin{vmatrix} \dot{y} & \dot{z} \\ y & z \end{vmatrix}\right) + \frac{4I_2}{(1+t^2)^2}\left(\dot{y} + \begin{vmatrix} \dot{z} & \dot{x} \\ z & x \end{vmatrix}\right)(-z) + \frac{4I_3}{(1+t^2)^2}\left(\dot{z} + \begin{vmatrix} \dot{x} & \dot{y} \\ x & y \end{vmatrix}\right)y, \qquad (101)$$

and so on. Writing it as

$$\frac{1+t^2}{2}p_x = \frac{2I_1}{1+t^2}\left(\dot{x} + \begin{vmatrix} \dot{y} & \dot{z} \\ y & z \end{vmatrix}\right) + \frac{2I_2}{1+t^2}(-z)\left(\dot{y} + \begin{vmatrix} \dot{z} & \dot{x} \\ z & x \end{vmatrix}\right) + \frac{2I_3}{1+t^2}y\left(\dot{z} + \begin{vmatrix} \dot{x} & \dot{y} \\ x & y \end{vmatrix}\right) \qquad (102)$$

and so on, and making use of (97), we find the set of equations (by cyclic permutation of the indices)

$$\frac{1+t^2}{2}p_x = I_1\omega_1 - zI_2\omega_2 + yI_3\omega_3 \qquad (103)$$

$$\frac{1+t^2}{2}p_y = zI_1\omega_1 + I_2\omega_2 - xI_3\omega_3 \qquad (104)$$

$$\frac{1+t^2}{2}p_z = -yI_1\omega_1 + xI_2\omega_2 + I_3\omega_3, \qquad (105)$$

or, defining

$$P_x \equiv \frac{1+t^2}{2} p_x, \text{etc} \quad (106)$$

and

$$\Omega_1 \equiv I_1 \omega_1, \text{etc}, \quad (107)$$

we finally get the system of three simultaneous equations for the three unknowns $\Omega i$

$$P_x = \Omega_1 - z\Omega_2 + y\Omega_3 \quad (108)$$
$$P_y = z\Omega_1 + \Omega_2 - x\Omega_3 \quad (109)$$
$$P_z = -y\Omega_1 + x\Omega_2 + \Omega_3 \quad (110)$$

Solving it by Cramer's method, we have for example

$$P_x = \Delta_1/\Delta, \quad (111)$$

where

$$\Delta \equiv \begin{vmatrix} 1 & -z & y \\ z & 1 & -x \\ -y & x & 1 \end{vmatrix}, \quad (112)$$

and

$$\Delta_1 = \begin{vmatrix} P_x & -z & y \\ P_y & 1 & -x \\ P_z & x & 1 \end{vmatrix}. \quad (113)$$

We find

$$\Delta = 1 + t^2, \quad (114)$$

and

$$\Delta_1 = P_x + (xP_x + yP_y + zP_z)x - \begin{vmatrix} y & z \\ P_y & P_z \end{vmatrix}. \qquad (115)$$

Thus, the result is

$$2I_1\omega_1 = p_x + (\vec{p} \cdot \vec{t})x + (\vec{p} \times \vec{t})_x, \qquad (116)$$

and so on, or finally (in compact form):

$$2II\vec{\omega} = \vec{p} + (\vec{p} \cdot \vec{t})\vec{t} + (\vec{p} \times \vec{t}). \qquad (117)$$

*iv) Spinors in orientation space continued*

Consider now the spin matrix (55), namely

$$S = s^\mu \sigma_\mu = \begin{pmatrix} s_3 & s_1 - is_2 \\ s_1 + is_2 & -s_3 \end{pmatrix} \qquad (118)$$

This can be thought of as an operator acting on 2×1 (column) matrices ψ, in the following manner:

$$\psi' = S\psi. \qquad (119)$$

We will prove that if ψ is a spinor, then so is ψ′ as well. In fact, S is transformed, by the action of a rotation transformation R, as follows:

$$\tilde{S} = RSR^*, \qquad (120)$$

where R* is the adjoint of R, with the property that

$$R^* = R^{-1}, \quad (121)$$

since R is unitary. On the other hand the spinor ψ is transformed according to

$$\tilde{\psi} = R\psi. \quad (122)$$

Thus the product Sψ is transformed to

$$RSR^{-1}R\psi = RS\psi. \quad (123)$$

But

$$RS\psi = R\psi', \quad (124)$$

which proves that ψ´ is a spinor as well.

Now let us find the norm of S. This is equal to

$$(s^\mu)^2(\sigma_\mu)^2 = I_1^2\omega_1^2(\sigma_x)^2 + I_2^2\omega_2^2(\sigma_y)^2 + I_3^2\omega_3^2(\sigma_z)^2. \quad (125)$$

But

$$\sigma_x^2 = \sigma_y^2 = \sigma_z^2 = 1, \quad (126)$$

so that the norm becomes equal to

$$I_1^2\omega_1^2 + I_2^2\omega_2^2 + I_3^2\omega_3^2 = s^2, \quad (127)$$

as expected. Note that this is equal to -detS, that is

$$\|S\| = -\det S. \quad (128)$$

v) *Spin is <u>not</u> angular momentum (even classically)*

Now, we change generalized coordinates from x, y, z to X, Y, Z defined by

$$\frac{2}{1+t^2}(\dot{\vec{t}} + \vec{t} \times \vec{t}) = \dot{\vec{X}} \qquad (129)$$

(cf. 97), that is

$$\frac{2}{1+x^2+y^2+z^2}\left(\dot{x} + \begin{vmatrix} \dot{y} & \dot{z} \\ y & z \end{vmatrix}\right) = \dot{X} \qquad (130)$$

etc. The variables X, Y, Z can then be found by integration, that is

$$X = 2\int \frac{dx}{1+x^2+y^2+z^2} + 2\int \frac{zdy - ydz}{1+x^2+y^2+z^2} \qquad (131)$$

etc. We have thus, for example,

$$X = 2I_x + 2zI_y - 2yI_z, \qquad (132)$$

where

$$I_x \equiv \int \frac{dx}{(1+y^2+z^2)+x^2} \equiv \int \frac{dx}{a_x^2+x^2} = \frac{1}{a_x}\arctan\frac{x}{a_x} \qquad (133)$$

(see Kappos, 1962), and in the same way

$$I_y \equiv \int \frac{dy}{(1+x^2+z^2)+y^2} \equiv \int \frac{dy}{a_y^2+y^2} = \frac{1}{a_y}\arctan\frac{y}{a_y} \qquad (134)$$

and

$$I_z \equiv \int \frac{dz}{(1+x^2+y^2)+z^2} \equiv \int \frac{dz}{a_z^2+z^2} = \frac{1}{a_z}\arctan\frac{z}{a_z}, \qquad (135)$$

so that

$$X = \frac{2}{a_x}\arctan\frac{x}{a_x} + \frac{2}{a_y}z\arctan\frac{y}{a_y} - \frac{2}{a_z}y\arctan\frac{z}{a_z}. \quad (136)$$

In the same way we find Y and Z.

By (97) and (99), we have

$$L = \tfrac{1}{2}I_1\dot{X}^2 + \tfrac{1}{2}I_2\dot{Y}^2 + \tfrac{1}{2}I_3\dot{Z}^2. \quad (137)$$

The new generalized momenta are

$$P_X = \frac{\partial L}{\partial \dot{X}} = I_1\dot{X} \quad (138)$$

etc, so that

$$\dot{X} = P_X/I_1 \quad (139)$$

etc. We have then

$$s_1 = I_1\omega_1 = I_1\dot{X} = P_X \quad (140)$$

etc.

Passing now to the quantum mechanical representation of quantities by operators, and because X and PX, etc, are conjugate quantities,

$$s_1 = P_X \rightarrow -i\hbar\frac{\partial}{\partial X} \quad (141)$$

etc. If we wanted to have correspondence of spin to angular momentum, we should have taken, instead of (129),

$$\omega_1 \equiv \frac{2}{1+x^2+y^2+z^2}\left(\dot{x}+\begin{vmatrix}\dot{y} & \dot{z}\\ y & z\end{vmatrix}\right)=\begin{vmatrix}\dot{\psi} & \dot{\zeta}\\ \psi & \zeta\end{vmatrix}, \quad (142)$$

etc, changing variables from x, y, z to χ, ψ, ζ. Then

$$L = \tfrac{1}{2} I_1 \begin{vmatrix}\dot{\psi} & \dot{\zeta}\\ \psi & \zeta\end{vmatrix}^2 + ..., \quad (143)$$

so that

$$P_\chi = \frac{\partial L}{\partial \dot{\chi}} = \tfrac{1}{2} I_2 \cdot 2 \begin{vmatrix}\dot{\zeta} & \dot{\chi}\\ \zeta & \chi\end{vmatrix}(-\zeta) + \tfrac{1}{2} I_3 \cdot 2 \begin{vmatrix}\dot{\chi} & \dot{\psi}\\ \chi & \psi\end{vmatrix}\psi, \quad (144)$$

etc. For the spin we will have

$$s_1 = I_1 \omega_1 = I_1 \begin{vmatrix}\dot{\psi} & \dot{\zeta}\\ \psi & \zeta\end{vmatrix}, \quad (145)$$

etc. We have to express thus ω1, ω2, ω3 as functions of Pχ, Pψ, Pζ. We obtain the following system of three algebraic simultaneous equations for the three unknowns ω1, ω2, ω3:

$$\left.\begin{aligned}P_\chi &= I_1 \cdot 0 & +I_2(-\zeta)\omega_2 & +I_3\psi\omega_3\\ P_\psi &= I_1 \zeta \omega_1 & +I_2 \cdot 0 & +I_3(-\chi)\omega_3\\ P_\zeta &= I_1(-\psi)\omega_1 & +I_2 \chi \omega_2 & +I_3 \cdot 0\end{aligned}\right\} \quad (146)$$

Solving it by Cramer's method we form the appropriate determinants and we find

$$\Delta = \begin{vmatrix} 0 & -\zeta I_2 & \psi I_3 \\ \zeta I_1 & 0 & -\chi I_3 \\ -\psi I_1 & \chi I_2 & 0 \end{vmatrix} = 0, \quad (147)$$

etc. We observe thus that the system has no solutions, which means that it is impossible to define

$$\Delta_1 = \begin{vmatrix} P_\chi & -\zeta I_2 & \psi I_3 \\ P_\psi & 0 & -\chi I_3 \\ P_\zeta & \chi I_2 & 0 \end{vmatrix} = \chi I_2 I_3 (\chi P_\chi + \psi P_\psi + \zeta P_\zeta), \quad (148)$$

variables χ, ψ, ζ by (142). Thus it is impossible to have correspondence of spin to angular momentum.

## 5. Quantum Mechanics in orientation space

*i) Representation of wave functions by spinors*

If we want to see the physical significance of the variables X, Y, Z, consider a rotation about the z-axis. Then, applying formulae (136), expressing the variables x, y, z by their definition (90)

$$x = \alpha t, \quad y = \beta t, \quad z = \gamma t, \quad (149)$$

substituting t also by its definition

$$t = \tan(\theta / 2), \quad (150)$$

and manipulating, we find

$$X = 0, \quad Y = 0, \quad Z = \theta. \quad (151)$$

Now, let us form the eigenvalue equation for the operator

$$\hat{s}_z = -i\hbar \frac{\partial}{\partial Z}. \quad (152)$$

We take as wave-function a scalar function ψ of (X, Y, Z) and we obtain

$$-i\hbar \frac{\partial \psi}{\partial Z} = \lambda \psi. \quad (153)$$

The solution of this differential equation is

$$\psi = e^{\frac{i}{\hbar}\lambda Z}. \quad (154)$$

If we consider the case (151), then

$$\psi = e^{\frac{i}{\hbar}\lambda\theta}. \quad (155)$$

But an orientation in orientation space is specified by a quaternion depending on θ/2, which therefore varies from 0 to 2π. Thus, θ in (155) varies from 0 to 4π. And since the exponential in (155) retains its original value if we change the argument in the exponent by 2kπ, we must demand

$$\frac{\lambda}{\hbar} \cdot 4\pi = 2k\pi, \quad (156)$$

where k is an integer. We obtain then for the eigenvalue λ

$$\lambda = k\frac{\hbar}{2}. \quad (157)$$

The eigenfunction (155), for the case k = 0, becomes

$$\psi = e^{i\theta/2}, \quad (158)$$

or

$$\psi = e^{i\theta'}, \quad (159)$$

with θ´ varying from 0 to 2π. After a rotation by θ = 2π, θ´ changes by π and ψ | θ´ε (π, 2π) is therefore the opposite function of ψ | θ´ ε (0, π). Since the quantity which changes sign by a rotation by θ = 2π is a spinor, we are led to conclude that the function ψ | θ ε (0, 4π) can be represented by a spinor function (ψ1, ψ2) | θ ε (0, 2π). In this way we have to represent rather the operator ŝz as a spin matrix. From now on thus, we will treat the spin as a spin matrix.

We have then

$$\hat{S} = -i\hbar \begin{pmatrix} \dfrac{\partial}{\partial Z} & \dfrac{\partial}{\partial X} - i\dfrac{\partial}{\partial Y} \\ \dfrac{\partial}{\partial X} + i\dfrac{\partial}{\partial Y} & -\dfrac{\partial}{\partial Z} \end{pmatrix} \quad (160)$$

and, in particular

$$\hat{S}_z = -i\hbar \begin{pmatrix} \dfrac{\partial}{\partial Z} & 0 \\ 0 & -\dfrac{\partial}{\partial Z} \end{pmatrix} \quad (161)$$

### ii) Quantum mechanical spin

The eigenvalue equation for the z-component of spin is thus

$$-i\hbar \begin{pmatrix} \dfrac{\partial}{\partial Z} & 0 \\ 0 & -\dfrac{\partial}{\partial Z} \end{pmatrix} \begin{pmatrix} \psi_1 \\ \psi_2 \end{pmatrix} = \lambda \begin{pmatrix} \psi_1 \\ \psi_2 \end{pmatrix} \quad (162)$$

so that we have

$$-i\hbar \dfrac{\partial \psi_1}{\partial Z} = \lambda \psi_1 \quad (163)$$

and

$$+i\hbar \dfrac{\partial \psi_2}{\partial Z} = \lambda \psi_2. \quad (164)$$

From (163) we find

$$\psi_1 = e^{\tfrac{i}{\hbar}\lambda \theta} \quad (165)$$

while from (164) we find

$$\Psi_2 = e^{-\frac{i}{\hbar}\lambda\theta} \quad (166)$$

with

$$\lambda = \pm \hbar/2 \quad (167)$$

in both cases, so that the eigen-spinors are

$$\begin{pmatrix} \Psi_1 \\ \Psi_2 \end{pmatrix}_+ = \begin{pmatrix} e^{i\theta/2} \\ e^{-i\theta/2} \end{pmatrix} \quad (168)$$

corresponding to the eigen-value $\lambda = +\hbar/2$, and

$$\begin{pmatrix} \Psi_1 \\ \Psi_2 \end{pmatrix}_- = \begin{pmatrix} e^{-i\theta/2} \\ e^{i\theta/2} \end{pmatrix} \quad (169)$$

corresponding to the eigen-value $\lambda = -\hbar/2$.

In the eigen-base, the eigen-spinors (168) and (169) will be, in terms of components, (1, 0) and (0, 1). Also in the eigen-base the operator $\hat{S}z$ will be a matrix ((a, c), (b, d)). Setting

$$\begin{pmatrix} a & b \\ c & d \end{pmatrix} \begin{pmatrix} 1 \\ 0 \end{pmatrix} = +\frac{\hbar}{2} \begin{pmatrix} 1 \\ 0 \end{pmatrix} \quad (170)$$

we find a = $\hbar/2$ and c = 0. Setting

$$\begin{pmatrix} a & b \\ c & d \end{pmatrix} \begin{pmatrix} 0 \\ 1 \end{pmatrix} = -\frac{\hbar}{2} \begin{pmatrix} 0 \\ 1 \end{pmatrix} \quad (171)$$

we find b = 0 and d = $-\hbar/2$. Thus the spin-matrix associated with the z-component of the spin, in terms of components, will be

$$\begin{pmatrix} a & b \\ c & d \end{pmatrix} = \frac{\hbar}{2}\begin{pmatrix} 1 & 0 \\ 0 & -1 \end{pmatrix} = \frac{\hbar}{2}\sigma_z. \quad (172)$$

In the same way we can find that for the x and y components we must have the matrices (ℏ/2)σx and (ℏ/2)σy. Ignoring the dependence on the "internal" (orientation) coordinates, we can thus form the eigenvalue equation for $\hat{S}z$ as

$$\frac{\hbar}{2}\begin{pmatrix} 1 & 0 \\ 0 & -1 \end{pmatrix}\begin{pmatrix} \psi_1 \\ \psi_2 \end{pmatrix} = \lambda \begin{pmatrix} \psi_1 \\ \psi_2 \end{pmatrix} \quad (173)$$

We have so

$$\frac{\hbar}{2}\psi_1 = \lambda\psi_1, \quad (174)$$

from where

$$\lambda = \hbar/2, \quad (175)$$

and

$$-\frac{\hbar}{2}\psi_2 = \lambda\psi_2, \quad (176)$$

from where

$$\lambda = -\hbar/2. \quad (177)$$

These equations are compatible if we take in fact as (normalized) eigen-spinors the spinor (1, 0), corresponding to λ = ℏ/2, and the spinor (0, 1), corresponding to λ = -ℏ/2.

*iii) The energy*

For the energy associated with the spin we have the Lagrangian

$$L = \frac{S_1^2}{2I_1} + \frac{S_2^2}{2I_2} + \frac{S_3^2}{2I_3}, \quad (178)$$

from which we take the Hamiltonian operator

$$\hat{H} = \frac{1}{2I_1}\left(\frac{\hbar}{2}\sigma_x\right)^2 + \ldots = \frac{\hbar^2}{8}\frac{3}{I}\sigma_0, \quad (179)$$

where we have set

$$\frac{3}{I} \equiv \frac{1}{I_1} + \frac{1}{I_2} + \frac{1}{I_3}. \quad (180)$$

Thus, the eigenvalue equation

$$\hat{H}\begin{pmatrix} a \\ b \end{pmatrix} = E\begin{pmatrix} a \\ b \end{pmatrix} \quad (181)$$

becomes

$$\frac{\hbar^2}{8}\frac{3}{I}\sigma_0\begin{pmatrix} a \\ b \end{pmatrix} = E\begin{pmatrix} a \\ b \end{pmatrix} \quad (182)$$

so that we find

$$E = \frac{\hbar^2}{8}\frac{3}{I}. \quad (183)$$

This energy is distributed equally in the three axes, because of symmetry, and we have for the z-axis, for example,

$$E = \frac{\hbar^2}{8I}. \quad (184)$$

On the other hand, for a rotation about the z-axis, we have the eigenvalue equation

$$\frac{\left(-i\hbar\frac{\partial}{\partial Z}\right)^2}{3I_3}\psi = E\psi, \quad (185)$$

or

$$-\frac{\hbar^2}{2I_3}\frac{\partial^2\psi}{\partial Z^2} = E\psi, \quad (186)$$

from which we find

$$\psi = e^{\pm i\frac{\sqrt{2I_3 E}}{\hbar}\theta}. \quad (187)$$

Then, from the condition

$$\frac{\sqrt{2I_3 E}}{\hbar} \cdot 4\pi = 2\pi, \quad (188)$$

we find

$$E = \frac{\hbar^2}{8I_3}. \quad (189)$$

Comparing (189) with (184), we conclude that

$$I_1 = I_2 = I_3 = I, \quad (190)$$

that is it is about a "spherical" particle.

*iv) Magnetic moment and determination of I*

A remark here about the magnetic moment of an electron in a magnetic field. Let l and m*l be the orbital angular momentum and magnetic moment respectively, and let s and m*s be the spin and the magnetic moment due to the spin respectively. We have

$$\frac{m^*_l}{l} = \frac{e}{2m}. \quad (191)$$

On the other hand, if one can show that in the same way

$$\frac{m^*_s}{s} = \frac{e}{2I}, \quad (192)$$

then, because from experiment (magnetic anomaly of the spin) we have

$$\frac{m^*_s}{s} = \frac{e}{m}, \quad (193)$$

we conclude by comparison that we must have

$$I = m/2. \quad (194)$$

*v) The total spin*

Let us now come to the total spin. We have

$$s^2 = s_1^2 + s_2^2 + s_3^2 = I_1^2 \omega_1^2 + ... \quad (195)$$

Passing to operators we have

$$\hat{s}^2 = \left(-i\hbar \frac{\partial}{\partial X} \sigma_x\right)^2 + ... = -\hbar^2 \frac{\partial^2}{\partial X^2} \sigma_0 + ..., \quad (196)$$

so that the eigenvalue equation is

$$\hat{s}^2\psi \equiv -\hbar^2 \frac{\partial^2 \psi}{\partial X^2} + ... = \lambda \psi. \quad (197)$$

Separating the variables, we set

$$\psi = \psi_X(X)\psi_Y(Y)\psi_Z(Z), \quad (198)$$

and we find

$$-\hbar^2 \frac{1}{\psi_X} \frac{\partial^2 \psi_X}{\partial X^2} = \lambda_X, ... \quad (\lambda_X + \lambda_Y + \lambda_Z = \lambda). \quad (199)$$

Solving we obtain

$$\psi_X(X) = e^{\pm i \frac{\sqrt{\lambda_X}}{\hbar} X}, ... \quad (200)$$

and from the conditions

$$\frac{\sqrt{\lambda_X}}{\hbar} \cdot 4\pi = 2k_X \pi, ... \quad (201)$$

we take finally

$$\lambda_X = k_X^2 \frac{\hbar^2}{4}, \quad (202)$$

Thus, for kX = kY = kZ = 1, we have for example

$$\lambda = \hbar^2/4 + \hbar^2/4 + \hbar^2/4 = \tfrac{1}{2}(\tfrac{1}{2}+1)\hbar^2. \quad (203)$$

## 6. Relativistic Mechanics in 10 dimensions

*i) Generalities*

We come now to the Special Theory of Relativity. Because now space and time are united in a four-dimensional space-time (four position coordinates), we must take as the orientation of a point (event) an arbitrary rotation of this four-dimensional space-time. This will depend on six parameters and will be given by a general Lorentz transformation, consisting of a spatial rotation plus a boost. Then the corresponding quaternion $L$ will be of the form of the well known quaternion R in ordinary three-dimensional space, but now expressing a rotation through a *complex* angle, say $\theta \mathbf{n} + i\boldsymbol{\alpha}$. We thus will have

$$L = \exp[(-\vec{\alpha} + i\theta\vec{n}) \cdot \vec{\sigma}/2]. \quad (204)$$

The conclusion is that the world will have a total of ten (10) dimensions: four (4) position dimensions (defined in space-time) plus six (6) orientation dimensions (defined in quaternion space). We will denote the 4 position coordinates by xi, defining the matrix (spinor)

$$X = x^i \sigma_i = \begin{pmatrix} t+z & x-iy \\ x+iy & t-z \end{pmatrix} \quad (205)$$

while the orientation coordinates are the 3 components of $\theta\mathbf{n}$ (describing the spatial rotation) plus the 3 components of $\boldsymbol{\alpha}$ (describing the boost), defining the quaternion $L$.

*ii) Transformations in position and in orientation space*

The transformation law for the position coordinates from one inertial frame to another sorresponding to a Lorentz spin matrix L (similar in structure to $L$) will be given by the well known formula

$$X' = LXL^*. \quad (206)$$

The transformation law for the orientation quaternion (depending on the orientation coordinates) will

be

$$L' = LL. \quad (207)$$

*iii) Reduction of L*

Write (204) in the form

$$L = \exp[i(\theta \vec{n} + i\alpha \vec{n}_\alpha) \cdot \vec{\sigma}/2] \quad (208)$$

and expand the exponential. We find

$$L = \sum_{p=0}^{\infty} \frac{1}{p!} [i(\theta \vec{n} + i\alpha \vec{n}_\alpha) \cdot \vec{\sigma}/2]^p. \quad (209)$$

We set for the complex vector $\theta \mathbf{n} + i\alpha \mathbf{n}\alpha$

$$\vec{N} \equiv (\theta \vec{n} + i\alpha \vec{n}_\alpha)/|\theta \vec{n} + i\alpha \vec{n}_\alpha| \equiv (\theta \vec{n} + i\alpha \vec{n}_\alpha)/N \quad (210)$$

Thus (209) becomes

$$L = \sum_{\text{even } p} \frac{1}{p!} (iN\vec{N} \cdot \vec{\sigma}/2)^p + \sum_{\text{odd } p} (iN\vec{N} \cdot \vec{\sigma}/2)^p. \quad (211)$$

Then, because of the identity

$$(\vec{a} \cdot \vec{\sigma})(\vec{b} \cdot \vec{\sigma}) = (\vec{a} \cdot \vec{b}) + i(\vec{a} \times \vec{b}) \cdot \vec{\sigma} \quad (212)$$

we have

$$(iN\vec{N} \cdot \vec{\sigma}/2)^2 = (iN\vec{N}/2)^2 = (iN/2)^2, \quad (213)$$

so that

$$\sum_{even\ p} \frac{1}{p!}(iN\vec{N}\cdot\vec{\sigma}/2)^p = \sum_{even\ p} \frac{1}{p!}(iN/2)^p = \cos(N/2) \qquad (214)$$

On the other hand

$$\sum_{odd\ p} \frac{1}{p!}(iN\vec{N}\cdot\vec{\sigma}/2)^p = (iN\vec{N}\cdot\vec{\sigma}/2)\sum_{odd\ p}\frac{1}{p!}(iN/2)^{p-1} =$$

$$= (\vec{N}\cdot\vec{\sigma})\sum_{odd\ p}\frac{1}{p!}(iN/2)^p = (\vec{N}\cdot\vec{\sigma})i\sin(N/2), \qquad (215)$$

so that (211) finally becomes

$$L = \cos(N/2) + i\sin(N/2)\vec{N}\cdot\vec{\sigma} \qquad (216)$$

*iv) The norm of L*

Concerning the norm ¶L¶ of L, we have

$$\|L\| = \cos^2(N/2) - \sin^2(N/2)N_x^2\sigma_x^2 - \sin^2(N/2)N_y^2\sigma_y^2 - \sin^2(N/2)N_z^2\sigma_z^2. \qquad (217)$$

But because

$$\sigma_x^2 = \sigma_y^2 = \sigma_z^2 = 1 \qquad (218)$$

and

$$N_x^2 + N_y^2 + N_z^2 = 1, \qquad (219)$$

we find

$$\|L\| = \cos^2(N/2) - \sin^2(N/2), \qquad (220)$$

from where finally

$$\|L\| = \cos N. \qquad (221)$$

For a pure spatial rotation ($\alpha = 0$), we have

$$\|L\| = \cos\theta, \qquad (222)$$

while for a pure boost ($\theta = 0$), we have

$$\|L\| = \cosh\alpha. \qquad (223)$$

v) *Pure spatial rotations*

In order to get an insight, consider a pure spatial rotation ($\alpha = 0$). As we have seen, the Lagrangian is

$$L = \frac{1}{2}I\omega^2. \qquad (224)$$

Since the rotation (for a free particle always) is uniform, we have for the *actual* action

$$S = \int \frac{1}{2}I\frac{\theta^2}{t^2}\,dt = \frac{1}{2}I\frac{\theta^2}{t^2}t = \frac{1}{2}I\frac{\theta^2}{t} \qquad (225)$$

Thus, the canonical momentum conjugate to $\theta$ is

$$p_\theta = \frac{\partial S}{\partial \theta} = \frac{1}{2}I\frac{2\theta}{t} = I\omega, \qquad (226)$$

that is right the spin s. Remembering that $I = m/2$, we can write, setting $m = 1$,

$$L = (1/4)\omega^2 \qquad (227)$$

and

$$s = (1/2)\omega. \qquad (228)$$

We could find the same things by another way. First of all the transition from one orientation to another can be performed by a series of successive infinitesimal rotations $\exp[i(d\theta/2)\mathbf{n}\cdot\boldsymbol{\sigma}]$. Since we know that infinitesimal rotations commute, the whole transition can be thought of as resulting from the rotation $\exp[i\int(d\theta/2)\mathbf{n}\cdot\boldsymbol{\sigma}]$. Now we demand that the integral of the norm of the integrand in the in the integral of the exponent is minimum for the actual path. This is an expression of Maupertuis´ principle. But this integral is obviously minimum when $\mathbf{n}$ remains constant during the transition. Then we simply sum up $d\theta$´s and the transition will be a rotation through an angle, say $\theta$, about a fixed axis, say $\mathbf{n}$, for the actual path. The norm of the exponent for this rotation $\exp[i(\theta/2)\mathbf{n}\cdot\boldsymbol{\sigma}]$ will be $(\theta^2)/4$.

Suppose that the rotation through the angle $\theta$ takes place in time t. If we divide t to small equal intervals $\Delta t$, and we denote the corresponding angles by $\Delta\theta$, we now demand that the sum $\Sigma\,[(\Delta\theta)^2]$ is a minimum. This is an expression of Hamilton´s principle. But this happens when all $(\Delta\theta)$´s are also equal to one another. Thus the rotation must be performed with constant angular velocity $\omega = (\Delta\theta)/(\Delta t) = \theta/t$. We now define the action by $S = (\theta^2)/(4t)$. Then we have that $S = [(\theta^2)/(4t^2)]t = (¼)(\omega^2)t$. But this must be the time integral of the Lagrangian. Hence we get for the Lagrangian formula (227). On the other hand the derivative of the action with respect to the coordinate gives the momentum. Hence formula (228).

### 7. Relativistic Quantum Mechanics in the 6-dimensional orientation space

*i) Pure spatial rotations*

We do not expect something essentially different from the casse of Classical Mechanics to occur. We thus will concentrate to

*ii) Pure boosts*

Generalize now for the case $\theta\mathbf{n} \to \theta\mathbf{n} + i\boldsymbol{\alpha}$. We will then have the same formulas (227) and (228) with now ω being complex. Specialize then for the case of pure boost (θ = 0). Then ω = id(α**n**α)/dt, that is the angular velocity is imaginary, and so is the spin. But the Lagrangian is real, namely L = = -(1/4)(dα/dt)^2. Consider a boost in the z-direction. Then **n**α = **k**, ω = idα/dt, s = (½)idα/dt, and L = -(1/4)(dα/dt)^2.

a. The spin

Since the spin is the canonical momentum conjugate to α, and considering "motion" in the z-direction, the spin operator will be

$$\hat{S} = -i\hbar \frac{\partial}{\partial \alpha} \qquad (229)$$

Thus if we denote by iA the (imaginary) eigenvalue, the eigenvalue equation takes the form

$$\hbar \psi' + A\psi = 0. \qquad (230)$$

The solution of this equation is

$$\psi = e^{-\frac{A}{\hbar}\alpha}. \qquad (231)$$

From the normalization condition

$$\int_0^\infty \psi \psi^* \, d\alpha = 1, \qquad (232)$$

if we substitute ψ in it, we obtain

$$\frac{\hbar}{2A} = 1, \qquad (233)$$

so that

$$A = \hbar/2, \qquad (234)$$

and the eigenfunction becomes

$$\psi = e^{-\alpha/2}. \quad (235)$$

Note that, equating the eigenvalue $iA = i\hbar/2$ with the spin $s = (½)id\alpha/dt$, we obtain

$$d\alpha/dt = \hbar, \quad (236)$$

and the Lagrangian $L = -(1/4)(d\alpha/dt)^2$ becomes

$$L = -(1/4)\hbar^2. \quad (237)$$

b. The energy

The Lagrangian is given, as we know, by

$$L = (½)I\omega^2, \quad (238)$$

with

$$\omega = d(i\alpha)/dt, \quad (239)$$

so that the action will be given by

$$S = (½)I(i\alpha)^2/t. \quad (240)$$

We thus find for the generalized momentum $p_{i\alpha}$, conjugate to the generalized coordinate $i\alpha$, with generalized velocity $\omega$,

$$p_{i\alpha} = Id(i\alpha)/dt, \quad (241)$$

or, calling it the *spin* s,

$$s = I\omega. \quad (242)$$

Thus the Lagrangian, in terms of s, becomes

$$L = \frac{1}{2I} s^2. \quad (243)$$

In Quantum Mechanics, as we already know, the spin operator is

$$\hat{s} = -i\hbar \frac{\partial}{\partial \alpha}, \quad (244)$$

so that the Lagrangian operator will be

$$\hat{L} = -\frac{1}{2I} \hbar^2 \frac{\partial^2}{\partial \alpha^2}. \quad (245)$$

In order to set up the Schrodinger equation, the Hamiltonian operator is needed rather than L. Thus we have to insert to our formulae a potential energy. This is just the "kinetic energy" known from Special Relativity,

$$U = \frac{m}{\sqrt{1-\beta^2}}, \quad (246)$$

since the velocity parameter β is related only to the *position* α by β = tanhα. We thus finally find U = = mcoshα, and the Hamiltonian operator will be

$$\hat{H} = -\frac{\hbar^2}{2I} \frac{\partial^2}{\partial \alpha^2} + m \cosh \alpha \quad (247)$$

Concerning I, it is equal to m/2. Thus, the Schrodinger equation will be

$$-\frac{\hbar^2}{m} \psi'' + (m \cosh \alpha - E)\psi = 0 \quad (E > U) \quad (248)$$

(see fig.2 for the energy diagram).

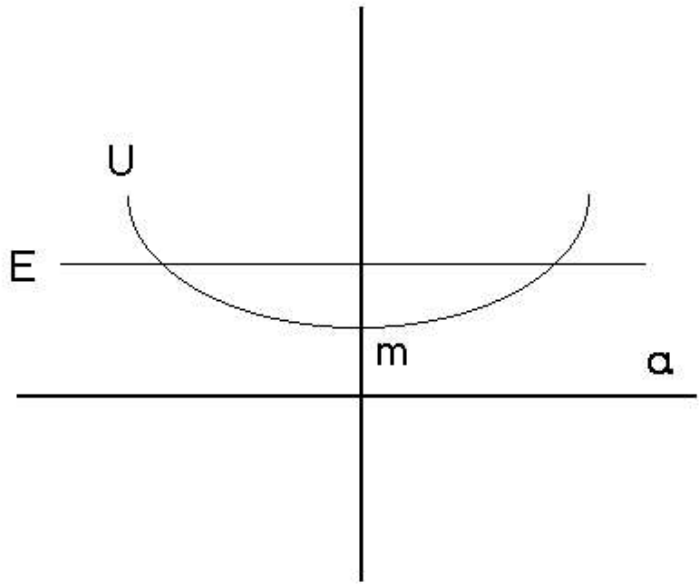

Fig. 2. The energy diagram associated with the Schrodinger equation (248).

The Schrodinger equation is of the form

$$y'' + f(x)y = 0. \quad (249)$$

In order to solve it, we set y = expu. Then y′ = u′expu, and y″ = (u″ + u′^2)expu. The equation then becomes

$$u'' + u'^2 + f(x) = 0, \quad (250)$$

and with the substitution u′ = z,

$$z' + z^2 + f(x) = 0. \quad (251)$$

This is of the Riccati type, and in particular

$$z' + z^2 - \frac{m}{\hbar^2}(m \cosh x - E) = 0, \quad (252)$$

which (in our case) cannot be solved by elementary methods (Kappos, 1966). But we can solve it approximately. If we set

$$\cosh x \cong 1 + x^2/2 \quad (253)$$

(Menzel, 1960), which means for small velocities, the Schrodinger equation takes the form

$$-\frac{2\hbar^2}{2m}\psi'' + \frac{m}{2}\alpha^2\psi = (E - m)\psi. \quad (254)$$

This is the case of the harmonic oscillator with ω = 1 & E → E – m (Matthews, 1963). Thus the eigenvalues of energy, we are primarily interested in, are given by the formula:

$$E_n = m + (½ + n)\hbar\sqrt{2} \quad (255)$$

c. A prediction

The above formula, written in the CGS system of units (c ≠ 1), becomes (see Wald, 1984)

$$E_n = mc^2 + (½ + n)\frac{\hbar\sqrt{2}}{\sec}\frac{c}{cm/\sec}. \quad (257)$$

These energy levels for plain electrons will give rise to a spectrum, which then will perhaps be detectable if the electrons are given the appropriate energy from outside. We find that this energy for successive-level photons is (see Allen, 1963)

$$\sqrt{2}\hbar c/cm \approx 2.8 \times 10^{-5} eV, \quad (257)$$

which corresponds to a frequency and wavelength

$$\nu \approx 6.75 GHz \,\&\, \lambda \approx 4.4 cm. \qquad (258)$$

Thus the corresponding spectral lines fall in the microwave part of the spectrum, and they can be detected in principle. If they are really detected, they will undoubtedly give a strong confirmation to the present theory.

**Note**

We rewrite formula (207) changing only the symbols to prevent confusion. Thus we write

$$A' = SA \quad \text{(a)}$$

where A (instead of $L$) is the orientation quaternion in orientation space, described by a Lorentz transformation, S (instead of L) is another Lorentz transformation, which transforms the coordinates, and A′ (instead of $L'$) is the transformed orientation A. Instead of taking this formula as expressing the transformation law for the orientations, it would be better to proceed as follows.

Following Jackson (*Classical Electrodynamics*), we make the ansatz A = expT, where T is an antisymmetric tensor. Then take A′ = expT′, where T′ = ST. In this way S need not be necessarily a Lorentz transformation. In the general case, that is in the General Theory of Relativity, S can be an arbitrary transformation applying to tensors, induced by an aritrary transformation in position space. In particular, S must be a Lorentz transformation only in the Special Theory of Relativity.

Thus, we would rather take as the formula describing the transformation law for orientations, instead of formula (a), the following formula:

$$A' = \exp\{S \ln A\} \quad \text{(b)}$$